\documentclass[runningheads]{llncs}
\usepackage[T1]{fontenc}
\usepackage{graphicx}
\usepackage{xcolor}
\usepackage{amssymb}

\newcommand\diff[1]{{{#1}}}
\newcommand\rem[1]{}

\begin{document}
\title{Whole Slide Image Classification of Salivary Gland Tumours}
%
%
\author{John Charlton \inst{1}\orcidID{0000-0001-8402-6723}, Ibrahim Alsanie \inst{2} \and
Syed Ali Khurram \inst{1}\orcidID{0000-0002-0378-9380}}
\authorrunning{J. Charlton, I. Alsanie, and S. A. Khurram.}
%
\institute{University of Sheffield, UK \email{j.charlton@sheffield.ac.uk}\\
\and King Saud University, Saudi Arabia
}
\maketitle              
\begin{abstract}
This work shows promising results using multiple instance learning on salivary gland tumours in classifying cancers on whole slide images. Utilising CTransPath as a patch-level feature extractor and CLAM as a feature aggregator, an F1 score of over 0.88 and AUROC of 0.92 are obtained for detecting cancer in whole slide images.




\keywords{Histology Image Classification \and Multiple Instance Learning  \and Salivary Gland Tumour.}
\end{abstract}
%
%





\section{Introduction}

Salivary gland tumours (SGTs) are a relatively rare group of heterogeneous neoplasms. These tumours represent approximately 3\% of all head and neck tumours \cite{gontarz_primary_2018,ito_salivary_2005,quixabeira_oliveira_epithelial_2023}. Artificial intelligence methods such as deep learning have been applied to many digital histological datasets \cite{gadermayr_multiple_2024,mahmood_artificial_2021} with very promising results. This includes high accuracy classification \cite{wang_transformer-based_2022} and segmentation \cite{raza_micro-net_2019} of numerous types of cancers.

Within the body of literature, there is a gap in knowledge regarding SGTs with applications using artificial intelligence. In particular, there is no work to the authors' knowledge that utilises the entirety of the whole slide image (WSI) in applying artificial intelligence to SGTs. Incorporating knowledge of the entire WSI is important for capturing large-scale histological and morphological information across the whole tissue.



To solve this issue, this work proposes a multiple instance learning (MIL) approach applied to WSIs of SGTs. This work classifies benign/malignant tumours, as well as classification of a particular type of malignant tumour (adenoid cystic carcinoma).
The work also compares the accuracy of the model when using two different feature extractors: ResNet-50 and CTransPath. It finds CTransPath to be the more accurate feature extractor, and predicts benign/malignant classification with an F1 score of 0.88 and AUROC of 0.92.




\section{Background and Methodology}

Multiple instance learning (MIL) \cite{carbonneau_multiple_2018,gadermayr_multiple_2024} is a variation on supervised learning. For MIL in this work, annotations are made at the WSI level (also known as the bag level in literature). 

Salivary gland tumours display a large amount of morphological diversity between tumour
types. This can be a challenge for models to accurately classify SGTs. In addition, the relative rarity of SGTs means datasets are difficult to obtain for use in training machine learning models.
Machine learning models have been successfully applied to SGTs at the patch level \cite{schulz_comparison_2023,prezioso_predictive_2022}, region of interest (ROI) scale \cite{alsanie_using_2022}, and using a graph-based approach \cite{alsanie_using_2023}. These works are able to classify SGTs with good accuracy, but they can be time-consuming and problematic for cancer subtyping, as high grade tumours are more challenging to annotate accurately.




Within this work two tasks were performed: benign/malignant classification, and adenoid cystic carcinoma/other classification. The first task was tested using two different feature extractors: ResNet-50 and CTransPath. The second task used only CTransPath as the feature extractor.

A dataset of 646 whole slide images of SGTs was used. Each WSI was labelled as either 'benign' (402 cases) or 'malignant' (242 cases). In addition, slides were categorised as adenoid cystic carcinoma (118 cases) or not (528 cases).
\diff{More images from other clinical groups will be included in future work to help test the model robustness across different clinical workflows. }

The workflow for these tasks was similar to other MIL approaches \cite{carbonneau_multiple_2018}. WSIs were split into smaller patches for feature extraction, then aggregated together utilising a feature aggregation model. For ResNet-50 feature extraction, the square patches were of side length 224 pixels and for CTransPath a patch was 256 pixels. Both ResNet-50 and CTransPath used the default weights of the model. CLAM was used for feature aggregation as was trained on the dataset.

Training the CLAM model was performed using k-fold validation for hyper-parameter tuning. A ratio of 80\%-10\%-10\% was used for training, validation, and testing respectively. \diff{k=10 folds were used, each data point appearing only once in the validation and once in the testing set.} 
\rem{This training process was repeated three times, once for ResNet-50 features for benign/malignant classification, once for CTranspath benign/malignant classification, and once for CTransPath adenoid cystic carcinoma classification.}


\section{Results}

\begin{figure}[htb]
\includegraphics[width=0.9\textwidth]{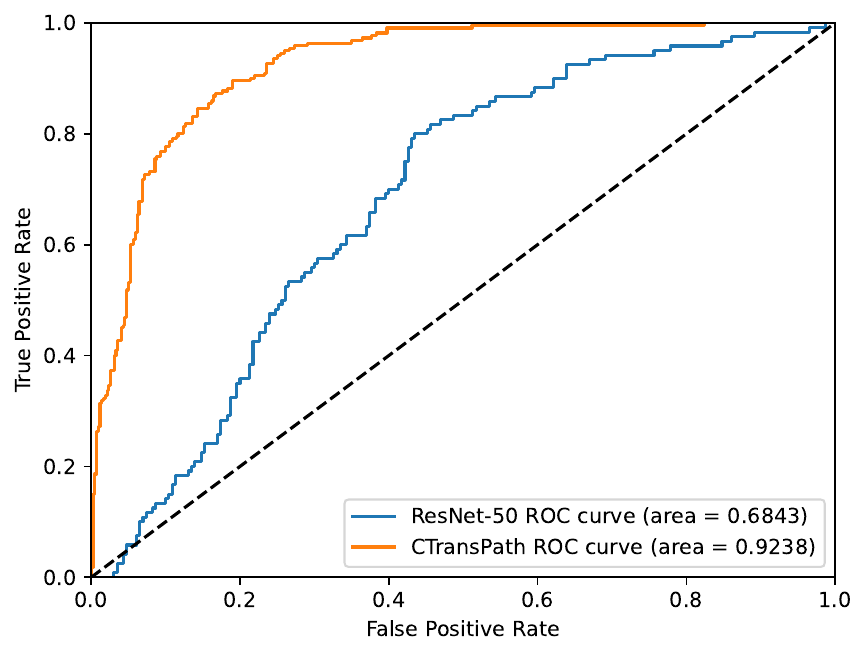}
\caption{ROC curve of Benign/Malignant classification. The blue ROC curve is for ResNet-50 features. The orange curve is for CTransPath features.} \label{fig:cancer}
\end{figure}

Figure \ref{fig:cancer} shows the two receiver operating characteristic (ROC) curves of binary classification of cancer. The blue curve is for features generated by ResNet-50. The orange curve is for features generated by CTransPath. It shows an area under the ROC (AUROC) of 0.92 for the method using CTransPath features, and 0.68 when using ResNet-50 features. 
\diff{For the CTransPath method the F1 score is 0.88, the precision is 0.90 and the recall is 0.88. The specificity is 0.92.
For the ResNet-50 method the F1 score is 0.72, the precision is 0.72, the recall is 0.77, and the specificity is 0.84.}

The figure shows higher accuracy when utilising CTransPath as the feature extractor compared to ResNet-50. This might be due to the datasets they were trained on. CTransPath was trained using histological images, and the features extracted by CTransPath appear to be more useful for this classification task.

The second task, Adenoid cystic carcinoma using CTransPath features with the CLAM feature aggregation model, achieved an AUROC of 0.96 and an F1 score of 0.84, displaying strong initial findings that a high grade SGT can be accurately classified for WSIs. \diff{It has a corresponding precision of 0.84, the recall is 0.77, and the specificity is 0.97.}

In conclusion, CTransPath features were found to provide greater accuracy in classification of cancer compared to ResNet-50 using a MIL approach. AUROCs of over 90\% were obtained for both tasks utilising CTransPath together with CLAM. \diff{The applicability of the model to other tasks is still to be explored, as well as more general conclusions about the comparison across more classification tasks.
Future work will compare against recent advancements of other architectures, including autoencoders and self-supervised learning to contextualise its performance.
}

\begin{figure}[htb]
\includegraphics[width=0.7\textwidth]{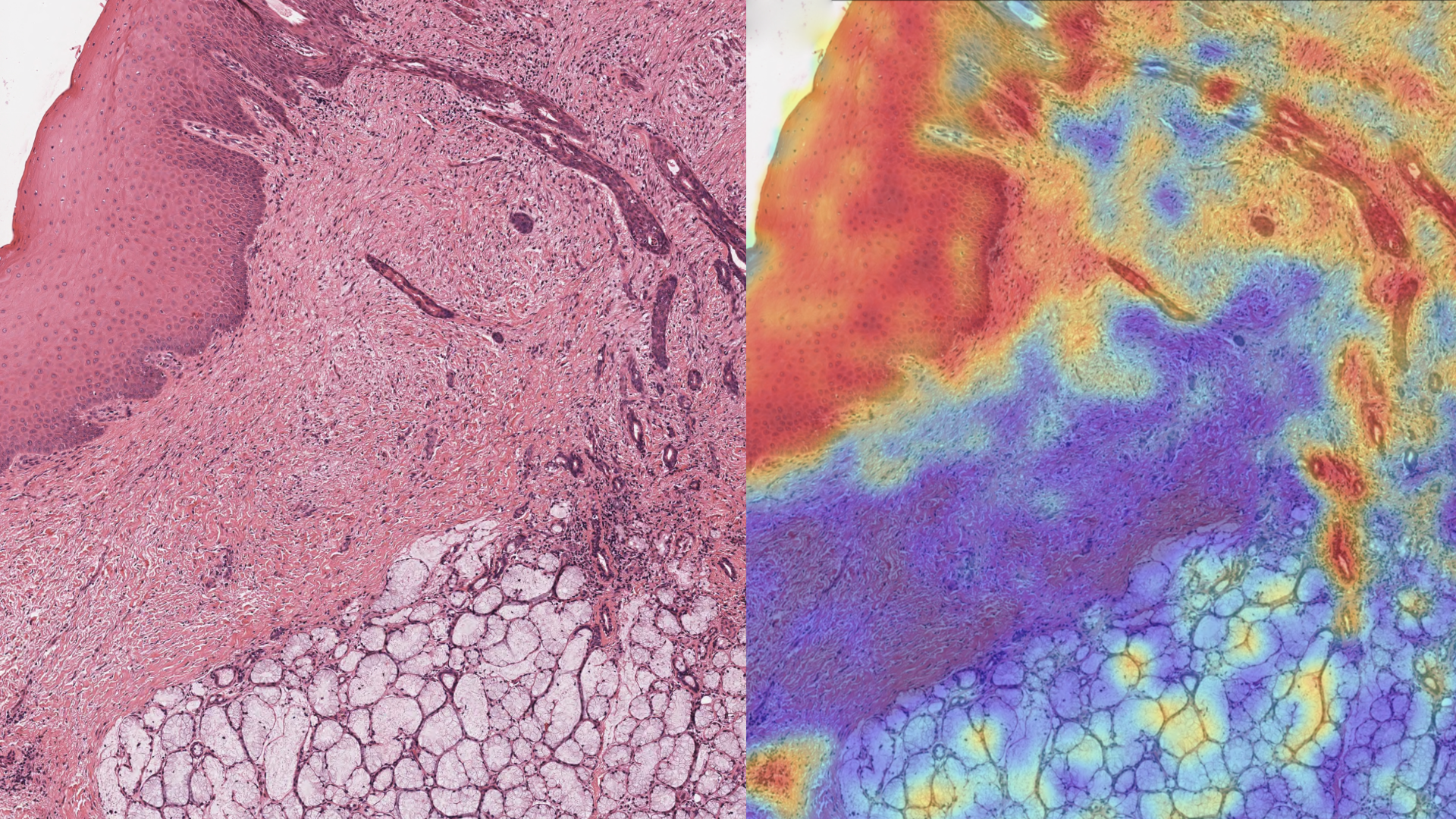}
\caption{Section of a whole slide image (WSI) showing the heatmap of attention. Areas highlighted in red are more important in deciding the categorisation of the whole image} \label{fig:wsi}
\end{figure}

The use of the attention mechanism in the CLAM model provides a focus for future study, as it highlights spatial regions within the WSI that are important for classification (see figure \ref{fig:wsi}). \diff{It attends differently between different tissue types, demonstrating its ability to account for pathological features.}
It follows that these regions are important in understanding the behaviour of cancer development within SGTs. This can be explored in future research to examine structural effects on important properties such as cancer behaviour, response to treatment, and patient survival.




\begin{credits}

\subsubsection{\discintname}
JC and IA have no competing interests to declare that are relevant to the content of this article. SAK is a shareholder in Histofy, an AI startup company.
\end{credits}
%
%
%
%
\bibliographystyle{splncs04}
\bibliography{references}





\end{document}